\documentstyle[prl,aps,twocolumn,epsf]{revtex}
\def\break#1{\pagebreak \vspace*{#1}}
\begin{document}


\title{Newton's Laws of Motion in Form of Riccati Equation}

\author{
Marek Nowakowski and Haret C. Rosu
}

\address{ 
Instituto de F\'{\i}sica de la Universidad de Guanajuato, Apdo Postal
E-143, Le\'on, Guanajuato, M\'exico}  

\maketitle
\widetext

\begin{abstract}
We discuss two applications of Riccati equation to Newton's laws of motion.
The first one is the motion of a particle under the influence of a power law
central potential $V(r)=k r^{\epsilon}$. For zero total energy we show 
that the equation of motion can be cast in the Riccati form.
We briefly show here an analogy to barotropic Friedmann-Robertson-Lemaitre
cosmology where the expansion of the 
universe can be also shown to obey a Riccati equation. A second application
in classical mechanics, where again the Riccati equation appears naturally,
are problems 
involving quadratic  friction. We use methods reminiscent to nonrelativistic 
supersymmetry to generalize and solve such problems. 

\end{abstract}
\vskip 0.1in

PACS number(s):  45.20.-d, 11.30 Pb 
\vskip 0.1in


\narrowtext

\section*{I. Introduction} 
It is known that Riccati equations, in general of the type

\begin{equation}
\frac{dy}{dx}=f(x)y^2+g(x)y +h(x)~,
\end{equation}
find surprisingly many applications in physics and mathematics. 
For example, supersymmetric quantum mechanics \cite{susy}, 
variational calculus \cite{zel}, nonlinear physics \cite{matv}, 
renormalization group equations for running coupling constants in 
quantum field theories \cite{odi} and thermodynamics \cite{rac}
are just a few topics where Riccati equations play 
a key role. 
The main reason for their ubiquity is that the change of function
\begin{equation}
y=-\frac{1}{f}\Big[\frac{d}{dx}\left(\log z\right) -\frac{g}{2}\Big]~,
\end{equation}
turns it into linear second-order differential equations of the form
\begin{equation}
\frac{d^2z}{dx^2}-\left(\frac{d}{dx}\log f \right)\frac{dz}{dx}-
\Big[\frac{g^2}{4}-\frac{1}{2}\frac{dg}{dx} +h -\frac{d}{dx}\log f\Big]z=0~,
\end{equation}
that stand as basic mathematical background for many areas of physics. 

Since the Riccati equation is a widely studied nonlinear equation, knowing that
the physical system under consideration can be brought into Riccati form has
certainly many advantages.

It is therefore of interest to look for yet different physical problems
which are governed by this first order nonlinear equation. This can be a 
starting point to new avenues in investigating analytical solutions of yet 
unsolved problems. In this paper we concentrate mainly on topics from 
classical mechanics and show that certain types of Newton's laws of motion are
equivalent to the Riccati equation. 


\break{1.17in}

\section*{II. THE POWER LAW CENTRAL POTENTIALS}
After implementation of the angular momentum conservation law, the equation
for the energy conservation in the case of a central potential $V(r)$ is 
given by the standard expression
\begin{equation} \label{P1}
E=
\frac{1}{2}m\dot{r}^2+\frac{l^2}{2mr^2}+V(r)~.
\end{equation}
Taking a derivative with respect to time of Eq.~(\ref{P1}) results into a
second fundamental equation of the form
\begin{equation} \label{P2}
m\ddot{r}-
\frac{l^2}{mr^3}+\frac{dV(r)}{dr}=0~.
\end{equation}
Specializing from now on to a power law potential  \cite{nieto} 
\begin{equation} \label{P3}
V(r)=kr^{\epsilon}~,
\end{equation}
where $k$ is the coupling constant and the exponent $\epsilon$ can be either
positive or negative, we obtain from (\ref{P2})
\begin{equation} \label{P4}
V(r)=-\frac{m\ddot{r}r}{\epsilon}+
\frac{l^2}{\epsilon mr^2}~.
\end{equation}
Inserting the last equation in (\ref{P1}) gives
\begin{equation} \label{P5}
\frac{1}{2}m \dot{r}^2+\left(\frac{1}{2}+\frac{1}{\epsilon}\right)
\frac{l}{mr^2}-\frac{m\ddot{r}r}{\epsilon}-E=0~.
\end{equation}
Under the assumption of $E=0$, this expression leads to a 
Riccati form. 
Obviously with $E=0$, we restrict ourselves to the case $k<0$. To explicitly 
derive from (\ref{P5}) the Riccati equation we pass (as it is costummary in 
central
potential problems) to an angle $\theta$ as a free variable (i.e., we consider
$r(\theta (t))$. With
\begin{equation} \label{P6}
\dot{\theta}=\frac{l}{mr^2}~, \qquad r^{'}\equiv \frac{dr}{d\theta}
\end{equation}
and introducing 
\begin{equation} \label{P7}
\omega=\frac{r^{'}}{r}~, 
\end{equation}
it can be readily shown, after some algebraic manipulations, that (\ref{P5}) 
reduces to
\begin{equation} \label{P8}
\omega ^{'}=\frac{\epsilon+2}{2}\omega ^2+\frac{\epsilon+2}{2}~.
\end{equation}
This is the Riccati equation for the motion of a particle in a central power
law potential assuming $E=0$. It is worth noting that no information about 
the coupling constant $k$ enters the Riccati equation (\ref{P8}).
Essentially what we have shown is that any solution of (\ref{P1}) will also 
satisfy (\ref{P8}). The inverse is not necessarily true and should be 
examined in detail.
Indeed, the coupling constant $k$ should be explicitly contained in the solution
for $r(\theta)$ (see below).

A special case which deserves to be briefly mentioned is $\epsilon =-2$. With 
this exponent, the choice $E=0$ is, in general, only possible if 
$\frac{l^2}{2m}+k<0$. Then directly from (\ref{P1}) we conclude 
that $\frac{r^{'}}{r}$ is a constant which, of course, is compatible with the 
Riccati equation (\ref{P8}). However, this constant cannot be determined by 
means of
(\ref{P8}). This feature is also inherent in the general case.

To discuss the case $\epsilon \neq 2$, we first solve the Riccati 
equation (\ref{P8}). The solution can be easily found to be 
\begin{equation} \label{P9}
\omega (\theta)=\tan \left(\frac{\epsilon+2}{2}\theta +\frac{\beta}{2}\right)
=\frac{\sin [(\epsilon+2)\theta +\beta]}{\cos [(\epsilon+2)\theta +\beta]
+1}~,
\end{equation}
where $\beta$ plays a role of the integration constant. Going back to the 
definition of $\omega$ in (\ref{P7}) we arrive at a solution for $r(\theta)$
\begin{equation} \label{P10}
r (\theta)=\frac{R}{[1+\cos((\epsilon +2)\theta +\beta)]^{1/(\epsilon +2)}}~,
\end{equation}
where $R$ is a constant. As in the case $\epsilon=-2$ this constant can only
be determined by inserting (\ref{P8}) into (\ref{P1}). The result is
\begin{equation} \label{P11}
R=\left(\frac{l^2}{m|k|}\right)^{\frac{1}{\epsilon+2}}~.
\end{equation}
The last two equations represent then the analytical solution of the posed 
problem. We obtained this solution by transforming the original problem into
a Riccati equation. It might be that the laws of motion in Riccati form are
only a curiosity. Given, however, the fact that only a few analytical solutions 
of the central potential problem are known, it is certainly a useful curiosity.
Furthermore, it is not excluded that this novel way opens new more general 
methods to solve problems in mechanics. In this context, we would like to 
mention 
here a yet different connection of the central potential problem with the 
Ermakov nonlinear differential equation \cite{er}. We refer to the 
following form of the latter equation 
\cite{esp} 
\begin{equation} \label{P12}
q(x)\frac{d^2y}{dx^2}+y(x)\frac{d^2q(x)}{dx^2}=\frac{1}{q^2(x)}
f\left(\frac{y}{q}\right)~,
\end{equation}
which can be solved by the integrals
\begin{equation} \label{P13}
\int \frac{dx}{q^2(x)}+a=\int \frac{d\left(\frac{x}{q}\right)}
{\sqrt{\phi\left(\frac{x}{q}\right)+b}}~,
\end{equation}
where $a$ and $b$ are integration constants and 
\begin{equation} \label{P14}
\phi(z)\equiv 2\int f(z)dz~.
\end{equation}
Taking $p={\rm const} =m$ and suitably rescaling the distance $r$ with the mass 
$m$, equation (\ref{P12}) is essentially identical to (\ref{P2}). 
Indeed, in this case the integrals in (\ref{P14}) give
\begin{equation} \label {P15}
t-t_0=\frac{1}{m}\int _{r_0}^{r}\frac{dr_{1}}{\sqrt{2mV(r_{1})-\frac{l^2}{r_{1}}
+b}}~,
\end{equation}
which, with a proper identification of $b$, is the same as directly integrating
(\ref{P2}). The interplay between the Ermakov equation and the central potential
problems can be a useful tool of studying both problems. We conjecture that
certain invariants of the Ermakov equation could be also applied to the 
central potential problems.

\section*{III. Cosmological Analogy} 
We want to point out here a beautiful but formal cosmological analogy to the 
results of the previous section.
We recall that in deriving the Riccati equation (9) we relied on a power law 
potential (\ref{P3}), a new parameter $\theta$ (the angle given in (\ref{P6})), and the 
assumption $E=0$. The analogy to cosmology is based on these observations.
In Friedmann-Robertson-Walker spacetime the set of Einstein's equations 
with the cosmological constant $\Lambda$ set to zero reduce to 
differential equations 
for the scale factor $a(t)$, which is a function of the comoving time $t$.
Together with the conservation of energy-momentum tensor they are given by
\begin{eqnarray}
3\ddot{a}(t)&=&-4\pi G(\rho +3p(\rho))a(t)~\\   
a(t)\ddot{a}(t)+2\dot{a}^{2}(t) +2\kappa &=& 4\pi G (\rho -p(\rho))a^2(t)\\
\dot{p}a^3(t)&=&\frac{d}{dt}\left(a^2(\rho +p(\rho)\right)~.
\end{eqnarray}
In the above $G$ is the Newtonian coupling constant, $p$ is the pressure, $\rho$
is the density and $\kappa$ can take the values $0,\pm 1$. Choosing the equation of 
state to be barotropic
\begin{equation} \label{C4}
p(\rho)=(\gamma -1)\rho ~,
\end{equation}
fixes essentially $\rho$ to obey a power law behaviour of the form
\begin{equation} \label{C5}
\rho=\rho _0 \left(\frac{a}{a_0}\right)^{-3\gamma} ~,
\end{equation}
and the remaining equation for $a(t)$ reduce to a single equation, viz
\begin{equation} \label {C6}
\frac{\ddot{a}(t)}{a(t)}+c\left(\frac{\dot{a}(t)}{a(t)}\right)^2
+c\frac{\kappa}{a^2(t)}=0 ~,\qquad c\equiv \frac{3}{2}c-1~.
\end{equation}
Introducing the conformal time $\eta$ by
\begin{equation} \label{C7}
\frac{d\eta}{dt}=\frac{1}{a(\eta)}~,
\end{equation}
it can be seen that (\ref{C6}) is equivalent to a Riccati equation in the 
function 
$u=\frac{a^{'}}{a}$, where the dot means derivation with respect to $\eta$
\begin{equation} \label{C8}
u^{'}+cu^2+\kappa c=0~.
\end{equation}
This cosmological Riccati equation has been previously obtained by Faraoni
\cite{far} and also discussed by Rosu \cite{ros}
in the context of late cosmological acceleration.
The formal analogy to the mechanical case is obvious: the 
condition $E=0$ corresponds to $\Lambda =0$, the angle $\theta$ is replaced by the 
conformal time $\eta$, and whereas in the mechanical example we had a power 
law behaviour of the potential, the barotropic equation of state forces 
upon $\rho$ to satisfy $\rho\propto a^{-3\gamma}$. As (\ref{P8}) does not 
contain 
the coupling constant $k$, the cosmological Riccati equation (\ref{C8}) loses 
the information about $G$.  

\section*{IV. Quadratic friction} 
Starting with a constant force $g$ (free fall, constant electric field, etc)
and adding a quadratic friction with a positive friction coefficient 
$\nu >0$, we have, per excellence, a Riccati equation for the Newton's law 
of motion
\begin{equation} \label{Q1}
\dot{v}=g^{'}-\alpha v^2~,
\end{equation}
with $g^{'}\equiv g/m$ and $\alpha \equiv \nu /m$. The general solution(which 
for reasons to be seen later in the text we denote by $v_p$) involves a free 
parameter $\lambda$ and reads \cite{dav}
\begin{equation} \label{Q2}
v_p(t;g^{'},\alpha , \lambda)=\frac{r}{\alpha}
\left(\frac{e^{rt}-\lambda e^{-rt}}{e^{rt}+\lambda e^{-rt}}\right)~,\qquad 
r\equiv \alpha g^{'}~.
\end{equation}
In the following we borrow some techniques from supersymmetric quantum 
mechanics. However, we do not follow strictly the supersymmetric scheme 
as the purposes in the quantum case and the mechanical case are quite different.
We define a new time-dependent force by
\begin{equation} \label{Q3}
\gamma(t;g^{'},\alpha , \lambda , \lambda _{1})\equiv
\dot{v}_{p}(t;g^{'},\alpha , \lambda)+\lambda _1
 v_{p}^{2}(t;g^{'},\alpha , \lambda)~.
\end{equation}
with a new parameter $\lambda _1 > 0$. We emphasize that (\ref{Q3}) is a 
definition given through the solution (\ref{Q2}). From (\ref{Q1}) it can be then 
deduced that the following equivalent form of $\gamma$ can be obtained
\begin{equation} \label{Q4}
\gamma = g^{'}-(\alpha -\lambda _1) v_{p}^{2}~.
\end{equation}
This resembles supersymmetric quantum mechanics and we might be tempted to 
compare $v_p$ to Witten's superpotential.

To the new force we again add a quadratic function with a friction coefficient
$\lambda _1$ such that the new equation of motion becomes
\begin{equation} \label{Q5}
\dot{v}=\gamma -\lambda _1 v^2~.
\end{equation}
This has the advantage that per construction $v_p$ is a particular solution of
(\ref{Q5}). Equipped with this fact, one can proceed to construct the general
solution which is a standard procedure in the general theory of the Riccati
equation. Before doing so, it is instructive to dwell upon the physical 
meaning of the new force $\gamma$. Imposing $g^{'}-
(\alpha -\lambda _1)\frac{r^2}{\alpha ^2}>0$, it can be seen that $\gamma >0$.
Moreover, as obvious from (\ref{Q2}) and (\ref{Q4}), $\gamma$ goes to a 
constant for large $t$ and has a kink-like behaviour. We can then envisage a 
situation where $\gamma$ is a `switch-on' function for a force becoming 
constant at some time.
As mentioned above, by construction the problem (\ref{Q5}) is solvable
because $v_p$ is a particular solution of (\ref{Q5}). By invoking the standard
Bernoulli ansatz for the general solution $v_g$, namely
\begin{equation} \label{Q6}
v_g=v_p+\frac{1}{V}~,
\end{equation}
we arrive at the differential equation (special case of the Bernoulli equation)
for $V$
\begin{equation} \label{Q7}
\dot{V}=2\lambda _1 v_p V+\lambda _1~.
\end{equation}
Writing $v_p$ as 
\begin{equation} \label{Q8}
v_p= -\frac{1}{\lambda _1} \frac{\dot{W}_p}{W_p}~,
\end{equation}
where
\begin{equation} \label{Q9}
W_p= e^{-\lambda _1 \int v_p dt}~,
\end{equation}
one is led to the solution for $V$
\begin{equation} \label{Q10}
V=\frac{\lambda _1 \int W_{p}^{2}dt +C}{W_{p}^{2}}~,
\end{equation}
The general solution is then given by
\begin{equation} \label{Q11}
v_g= v_p+ \frac{W_{p}^{2}}{\lambda _1 \int W_{p}^{2}dt +C}~.          
\end{equation}
The initial value problem, $v(0)=v_0$, is solved by fixing $C$ through
\begin{equation} \label{Q12}
v_0- \frac{r}{\alpha}\left( \frac{1-\lambda}{1+\lambda}\right)=\frac{1}{C}~.          
\end{equation}
Up to integrals, the equation of motion (\ref{Q5}) is solved. 
Setting $\lambda=e^{-2\delta}$, we can rewrite (\ref{Q2}) in the more 
convenient form
\begin{equation} \label{Q13}
v_p= \frac{r}{\alpha}{\rm tanh}(rt+\delta)~.          
\end{equation}
Then $W_p$ can be computed explicitly
\begin{equation} \label{Q14}
W_p=\frac{1}{[{\rm cosh}(rt+\delta)]^{\lambda _1/\alpha}}~.          
\end{equation}
It suffices to assume $\lambda _1=n\alpha, \, n\in N$ leading to integrals of 
the type $\int {\rm cosh}^{-n}(x)dx$, which can be solved in a closed analytical
form by recursion formulae.
Of course, the procedure outlined here can be generalized by starting with 
more complicated forces instead of the constant one.

\section*{IV. Conclusion} 
In this paper we have pointed out the usefulness of the Riccati equation in 
studying certain mechanical problems. We derived a Riccati equation for a
central potential problem of the power law type assuming $E=0$. This led us
to an analytical solution of the problem. In a second step, we generalized
the system of a constant force plus a quadratic friction to a time-dependent
force and friction. We argued that this time-dependent force serves 
as a `switch-on' function. The problem turned out to be solvable by means of a 
construction similar to supersymmetric quantum mechanics. As indicated in 
the text, both applications can be generalized.

\section*{Acknowledgment}
The first author thanks CONACyT for financial support 
through a Catedra Patrimonial fellowship. 



\newpage




\end{document}